# First photometric investigation of some ASAS binary systems

Burak ULAŞ

İzmir Türk College Planetarium, İzmir, TURKEY

burak.ulas@itk.k12.tr

**Abstract:** We present the first light curve solutions of the four binary systems observed and catalogued by the All Sky Automated Survey program. The light curves are analysed by using PHOEBE software, which is based on the Wilson-Devinney method. The light curve parameters are derived and the estimated physical parameters are calculated from the result of the analyses. We compare our targets to well-known binaries of similar type on the Hertzsprung-Russell diagram and mass-radius plane. The evolutionary states of the components are also discussed.

1. Introduction

ASAS (All Sky Automated Survey) [1] was set to monitor all sky for the stars brighter than $14^m$. The project focuses on the photometric variations of selected targets and aims the classification of the variable stars. The two active telescopes of the project are located in Las Campanas Observatory, Chile and Hawaii, USA. The survey is carried on with fully automated equipment with *V* and *I* filters and heretofore more than 15 million light curves were catalogued.

Binary stars are important tools to understand the stellar lifecycles in the universe. Since most stars are binary stars, the binary star studies also allow us to determine the observed universe in terms of stellar evolution and distribution. Semi-detached binary systems, in particular, are important in understanding the observed mass exchanges between the components and their effects on the evolutionary properties of the binary stars. They also allow us to reveal the role of binarity in the certain phases of the evolution of the stars. Therefore, the studies focusing on the derivation of the absolute physical properties and the evolution of the binary systems are crucial to figuring out the remarkable part of the observable properties of the stars.

Our study is the first detailed investigation of the four binary systems. The information about our targets in literature is extremely limited. [2] listed ASAS 070530+1521.0 in their red and infrared magnitude tables. The target was also catalogued by [3] based on photographs stored at the Warner and Swasey Observatory. [4] listed the radial velocity data of ASAS 163424-4919.5 in their survey and provided the $B_T$, $V_T$, J, H, and K magnitudes of the system. The positional and photometric properties of the systems are given in Table 1 and Table 2. The effect of the interstellar medium to the light of the systems was derived by using the interstellar reddening distribution study by [9]. Considering the calibration on the diagrams and maps in the paper, the galactic coordinates and distances of our targets correspond that the mean weighted colour excess $E_y$ was estimated as $0^m.04$ for A070, $0^m.17$ for A073, between $0^m.1$ and $0^m.2$ for A143, $0^m.19$ for A163.

In the next section, the properties of the light curve data, effective temperature determination and the analyses of the light curves of four binary stars are presented in detail. The third section deal with the results and the location of the components in the HR Diagram and mass-radius plane. A brief discussion on the evolutionary status of the systems is made in the last section.

**2. Materials and methods**

**2.1. Properties of the data**

The light curves of the targets were obtained in *V* passband provided from ASAS database[1] (Figure 1). They consist of 352, 668, 632 and 567 data points for A070, A073, A143, A163, respectively. The datasets cover 2544 days for A070, 3298 days for A073, 3189 days for A143 and 3180 days for A163.

**2.2. Estimation of the effective temperatures**

During the light curve analyses of binary stars, the accurate assumption of effective temperatures is critical, in order to reveal more realistic results. Therefore, since our targets are not the well-studied sources we made different efforts to achieve the proper assumptions for the effective temperatures of primary components. The spectral type of the system A070, F8V, was given by [3], therefore, the temperature value to be used in the analysis was estimated from the calibrations given by [10]. The lack of magnitudes obtained during the maximum phases of the light curves of A073 and A143 in literature directed us to use the effective temperature value given by Gaia DR2 [5, 6]. The values are 6368 K and 5861 K for A073 and A143, respectively.

We used $(V-K)_0$ intrinsic colour to estimate the temperature value for the target A163. The process was done in five steps: *(i)* The *V* magnitude at maximum brightness were taken from the ASAS database. *(ii)* The *K* magnitude value was adopted from the 2MASS All-Sky Catalogue of Point Sources [11] by checking the phase of the binary system at the observation time given in the catalogue. According to the catalogue, the 2MASS observation made during phase 0.79 for A163. Therefore, we ensured that the *K* magnitude was obtained at the neighbouring phase of the system's maximum light. *(iii)* To calculate the intrinsic *(V-K)₀* value we used the extinction ratio, *k,* given in Table 1 of [12]. The authors proposed an equation for extinction ratio:

$$k = \frac{E(colour)}{E(B-V)} \qquad (1)$$

where *E(colour)* is the extinction in any colour. It is obvious that the calculation of the *E(V – K)* needs estimating the *E(B – V)* colour. Thus, we constructed a Spectral Energy Distribution (SED) for the target from the available photometric data in the Vizier database [13]. The photometric data were fit by using parameter-grid search in order to obtain the optimized Kurucz atmosphere model [14]. Then *E(B – V)* value, 0.147, were calculated for A163 from the difference between reddened and dereddened models. The fit of distribution with the reddened model is shown in Figure 2. A wider explanation on the method of reddening calculation by fitting the SED can be found in [15]. *(iv)* Equation (1) takes the following form for the *(V – K)* colour [12]:

$$k = \frac{E(V-K)}{E(B-V)} = 2.71. \qquad (2)$$

---
[1] www.astrouw.edu.pl/asas

Therefore, the colour excess *E(V – K)* was determined and the *(V – K)₀* was easily computed by using the relation *(V – K)₀ = (V – K) – E(V – K)*. *(v)* Finally, we estimated the temperature from the Table 11 of [12] by using the *(V – K)₀* intrinsic value and by assuming that the star is a main sequence star with solar abundance. The estimated temperature value was used as fixed parameter during the analysis of the light curve of the system A163.

**2.3. Analyses of the Light Curves**

The light curves analysed separately by using PHOEBE [16] software which uses Wilson-Devinney method [17]. The program employs the best fit using differential corrections to derive the most appropriate parameters. The systems are classified as semi-detached binaries in ASAS webpage. However, since there is no any other reliable reference on the geometrical configuration of the systems in the literature we started to analyse the light curves by assuming that the systems are detached binaries. In each step of the analyses, we examined the surface potential values of the components in case of they exceed inner or outer critical potential value. The physically meaningful results for the systems A070 and A073 were achieved by the assumption of detached configuration, viz. Mode 2 of the code. In a certain step of the analysis of A143 in detached binary mode, we encountered a warning message in the output file of the program stating that the secondary component exceeded critical lobe. Therefore, we reanalysed the light curve in Mode 5 (semi-detached binary, secondary star fills Roche lobe) and yielded physically reasonable results. The most challenging series of analyses were that of A163. We started the analysis in Mode 2 which resulted that the primary component exceeded the outer contact surface and directed us to Mode4 (semi-detached binary, primary star fills Roche lobe). However, a few runs of the program in Mode 4 stated that the secondary component surpasses the critical lobe. The Mode 3 (overcontact binary not in thermal contact) of the program was also employed and the code started to give unphysical results for the effective temperature of the secondary component. Finally, the physically meaningful results were achieved in Mode 5 (semi-detached binary, secondary star fills Roche lobe).

Since there is no any mass ratio ($q$) value for the systems published in the literature we first applied the q-search technique to the light curves of the systems by using the appropriate modes mentioned in the previous paragraph in order to derive the appropriate initial mass ratio values. The results of the q-search are shown in Figure 3. We used the q values which correspond to the minimum $\chi^2$ as the initial mass ratio values for our further analyses.

The analyses were applied to the normalized data points. Therefore, before the analyses, following steps were pursued during the normalization process. We first removed the data points labelled with C and D letters since they are remarked as improper or useless in the header of the data files. Then, we calculated the phase values for each data by using the ephermides given by ASAS webpage. Finally, we generated normalized data points by extracting data points smaller than 1σ and using the average for phase interval $\Delta\varphi = 0.04$. Consequently, we applied the analyses on 25 normalized data points yielded from the data of each target.

The free parameters of the final analyses in Mode 5 were the inclination $i$, mass ratio $q$, the temperature of the secondary component $T_2$, surface potential of the primary $\Omega_1$ and luminosity of the primary component $L_1$. The surface potential of the secondary component $\Omega_2$ was also set free during the solutions in Mode 2. The *(B – V)* colours of the binary systems listed in Table 2 indicate that both or, at least, one of their components should have convective envelopes because the granulation boundary for the main sequence stars show up approximately in the spectral type F0 [18] which corresponds to *(B – V)* = 0.3 [10]. Therefore, the gravity darkening coefficients $g_1$ and $g_2$ were calculated as in [19] and the albedo values $A_1$ and $A_2$ were adopted from [20] by assuming that the components have convective envelopes. The results of the analyses are listed in Table 3. The comparison between the calculated light curves and observations were given in Figure 4.

Since the scattering in the light curves may obscure a possible O'Connell effect we investigated the data and inquired the trend of out of eclipse parts of the curves. The characteristic trace of the effect is that one of the maxima of the light curve is brighter than the other. The reason that causes the O'Connell effect can be various: asymmetrically distributed starspots, circumstellar dust and gas or a hotspot caused by mass transfer [21]. To investigate the effect, we plotted the magnitudes at the maximum phases of the light curves versus cycle numbers (Figure 5). During the process, we limit the range within 0.2-0.3 for first quadrature and 0.7-0.8 for second quadrature. We compared the data to straight lines which intersect the magnitude axes at the maximum value ($9^m.83$ for A070, $11^m.73$ for A073 and A143, $11^m.51$ for A163) that was used for the calculation of the fluxes during light curve analyses. This allowed us to observe the variations in magnitude values from the lines with cycle number. The results indicated that it is not possible to mention notable changes in the lights and obvious difference between two maxima that can be attributed to O'Connell effect.

It is also worth to point out that the fill-out factors, $f = \frac{\Omega_{cr}}{\Omega} - 1$, [22] for A143 and A163 were found to be -0.01 and -0.05, respectively. These values are very close to that of typical near-contact binary systems [22]. It may be hypothesized that the two systems are near contact binaries and they can be considered as a member of FO Virginis subclass according to the classification of [23] since a remarkable difference between two maxima of the light curves did not detected and the secondary components are at their Roche lobe, namely the surface potential values of them are equal to potential value of the inner contact surface.

3. **Results**

The light curve solutions of four binary systems are presented. Our analyses are the first analyses of these systems in the literature. Based on the analyses, we determined the estimated absolute parameters of the systems by using the effective temperature-spectral type-mass calibration of [10] and listed them in Table 4. In the determination process we calculated the estimated parameter values by applying interpolation between two neighbouring data given by [10]. The geometric models of the components of our targets are also plotted by using the obtained parameters in Table 3 and shown in Figure 6. These geometric configurations form characteristic detached and semi-detached light curves (Figure 4) which show a deep (primary minimum) and a shallow (secondary minimum) minima. The deeper minimum is observed since the cooler secondary occult the hotter component about phase 0.0 and hence it causes a decrement in the observed flux value. The shallower minimum, on the other hand, occurs when the cooler component eclipsed by the hotter one.

Our investigation also covers the comparison of the systems to other systems of similar types. We compare our detached targets, A070 and A073, to 162 detached binaries from [24] and semi-detached systems, A143 and A163, to 61 semi-detached Algol type binaries from [25] on the Hertzsprung-Russell diagram and the mass-radius plane (Figure 7). According to our estimated absolute parameters derived from the analyses and calibrations, it can be mentioned that the secondary components of A070 and A073 show discrepancy from the gathering of the systems of same type on mass-radius plane (upper left panel of Figure 7) while they are in good agreement in the Hertzsprung-Russel diagram (upper right panel of Figure 7). The radii of the primary components of semi-detached systems, A143 and A163, seem larger than expected comparing to primaries with the same mass (lower left panel of Figure 7) and the secondaries are underluminous than the others in the Hertzsprung-Russel diagram (lower right panel of Figure 7).

**4.    Discussion**

Since our systems are found to be detached (A070 and A073) and semi-detached systems (A143 and A163) according to our results from the analyses and the calibrations it is opportune to underline the evolutionary link between these two types of systems: The progenitor of a semi-detached binary is a detached system. In a detached system, the more massive component fills its Roche lobe due to its faster evolution. The Roche lobe filling causes the mass transfer to the originally less massive (present primary of semi-detached system) component [26]. The mass transfer causes change in the mass ratio and a semi-detached system forms when the mass ratio reversal occurs.

We represented the location of our targets with the evolutionary tracks of some binary systems in Figure 8, in order to estimate the evolutionary paths of the systems. These tracks were created by using BSE: Binary Star Evolution code [27, 28]. During the derivation of the evolutionary lines, we set the initial values as follows: eccentricity $e_i$, between 0 and 1, mass of the primary component $M_{1i}$, between 1.0 $M_\odot$ and 10.0 $M_\odot$, mass of the secondary component $M_{2i}$, between 0.1 $M_\odot$ and $M_1$, and the orbital period of the system $P_i$, between 3 and 7 days. We also limit the maximum evolution time to 15 Gyr. All of stars in this study were assumed in solar abundances since they located very near the galactic disk plane as seen from Table 1. In Figure 8, the closest track passing through the location of the components among the thousands of results yielded by running the code was plotted.

Our models for semi-detached systems (c and d in Figure 8) show that the evolutionary tracks of the initially less massive components head towards the locations of present more massive components. That is, the current positions of the primaries are on the evolutionary tracks of the initially secondary components. This result also affirms the theory of the binary star evolution we mentioned in the first paragraph of the section.

We concluded that four targets investigated in this study are detached and semi-detached binaries according to our analyses. The need for spectroscopic observations is crucial for solving the photometric and spectroscopic data simultaneously, and deriving more sensitive parameters for the systems in question. More precise values for the parameters then allow us to estimate the more accurate scenarios for the evolution of the systems.

**Acknowledgments**

This research has made use of the SIMBAD database and the VizieR catalogue access tool, operated at CDS, Strasbourg, France. This work has made use of data from the European Space Agency (ESA) mission Gaia (https://www.cosmos.esa.int/gaia), processed by the Gaia Data Processing and Analysis Consortium (DPAC, https://www.cosmos.esa.int/web/gaia/dpac/consortium). Funding for the DPAC has been provided by national institutions, in particular the institutions participating in the Gaia Multilateral Agreement. This publication makes use of data products from the Two Micron All Sky Survey, which is a joint project of the University of Massachusetts and the Infrared Processing and Analysis Center/California Institute of Technology, funded by the National Aeronautics and Space Administration and the National Science Foundation. The author would like to thank the anonymous reviewers for their constructive comments and suggestions.

**Table 1.** Positional properties of the systems. RA and DEC stand for equatorial coordinates while l and b are the galactic coordinates. All coordinates are given in J2000 epoch[2]. Parallaxes are taken from Gaia DR2 [5,6].

| ID | ASAS ID | RA (h:m:s) | DEC (°:´:´´) | l (°) | b (°) | Parallax (mas) |
|---|---|---|---|---|---|---|
| **A070** | 070530+1521.0 | 07:05:30 | +15:20:56 | 200.702 | +9.980 | 2.4571 |
| **A073** | 073404-3215.9 | 07:34:04 | -32:15:54 | 246.258 | -5.984 | 1.4397 |
| **A143** | 143405-5814.5 | 14:34:04 | -58:14:32 | 316.114 | +1.978 | 0.6548 |
| **A163** | 163424-4919.5 | 16:34:24 | -49:19:30 | 335.533 | -1.105 | 1.5175 |

---

[2] simbad.u-strasbg.fr/simbad/

**Table 2.** Photometric properties of the systems. $T_0$ and P are listed by following the values given in ASAS web page and they indicate the times of primary minimum and the orbital period, respectively. $B_T$ and $V_T$ magnitudes are taken from Tycho-2 catalogue [7] while V and K magnitudes are adopted from ASAS web page and [8], respectively.

| ID | GAIA DR2 ID | TYCHO ID | $B_T$-$V_T$ (mag) | V-K (mag) | $T_0$ (days) | P (days) |
|---|---|---|---|---|---|---|
| A070 | 3359976628178895616 | 1345-1422-1 | 0.50 | 1.22 | 2452622.1 | 1.41042 |
| A073 | 5592279626810866816 | 7109-2944-1 | 0.68 | 1.32 | 2451904.2 | 1.554567 |
| A143 | 5891351393803839360 | 8691-1308-1 | 0.31 | 1.55 | 2451904.2 | 1.554567 |
| A163 | 5940517396305363968 | 8333-1693-1 | 0.32 | 1.47 | 2451930.9 | 1.007775 |

**Table 3.** The results of the light curve analyses. The standard errors in the last digit are given in parentheses for mass ratio, effective temperature, surface potential, luminosity and fractional radius. See text for the detailed explanation for the parameters.

| Parameter | A070 | A073 | A143 | A163 |
|---|---|---|---|---|
| i(°) | 74.6(1.9) | 77.1(2.3) | 82.9(1.8) | 87.5(1.8) |
| q | 0.23(3) | 0.26(3) | 0.33(1) | 0.139(7) |
| $T_1$ (K) | 6135 | 6368 | 5861 | 6540 |
| $T_2$ (K) | 5129(138) | 4458(212) | 3925(130) | 4771(139) |
| $\Omega_1$ | 2.5(2) | 2.6(1) | 2.57(2) | 2.21(3) |
| $\Omega_2$ | 2.8(2) | 2.9(2) | $\Omega_{cr}$=2.53 | $\Omega_{cr}$=2.08 |
| $A_{1,2}$ | 0.5 | 0.5 | 0.5 | 0.5 |
| $g_{1,2}$ | 0.32 | 0.32 | 0.32 | 0.32 |
| $\frac{L_1}{L_1+L_2}$ | 0.952(3) | 0.98(1) | 0.98(2) | 0.96(1) |
| $r_1$ | 0.441(3) | 0.44(3) | 0.469(9) | 0.51(1) |
| $r_2$ | 0.16(6) | 0.16(5) | 0.286(4) | 0.225(5) |

**Table 4.** The estimated absolute parameters of the systems. *M, R, T, L* and *a* corresponds to the mass (in solar unit), radius (in solar unit), effective temperature, luminosity (in solar unit) and the semi-major axis (in solar radius), respectively. 1 and 2 indices refer to primary (more massive) and secondary components. The effective temperature of the Sun is set to 5777 K [10] during the calculations and the standard errors in the last digit are given in parentheses for mass, radius, effective temperature, semi-major axis and luminosities of A143 and A163.

| Parameter | **A070** | **A073** | **A143** | **A163** |
|---|---|---|---|---|
| $M_1$ ($M_\odot$) | 0.9 | 1.54 | 1.02 | 1.3 |
| $M_2$ ($M_\odot$) | 0.21(3) | 0.40(5) | 0.34(1) | 0.181(9) |
| $R_1$ ($R_\odot$) | 2.5(2) | 2.4(2) | 3.01(6) | 2.51(5) |
| $R_2$ ($R_\odot$) | 0.9(7) | 0.9(5) | 1.74(4) | 1.05(4) |
| $T_1$ (K) | 6135 | 6368 | 5861 | 6540 |
| $T_2$ (K) | 5129(138) | 4458(212) | 3925(130) | 4771(139) |
| $L_1$ ($L_\odot$) | 8.1(1.2) | 8.4(1.2) | 9.5(4) | 10.2(4) |
| $L_2$ ($L_\odot$) | 0.5(7) | 0.4(3) | 0.64(3) | 0.51(4) |
| $a$ ($R_\odot$) | 5.6(1) | 5.4(1) | 6.40(5) | 4.93(3) |

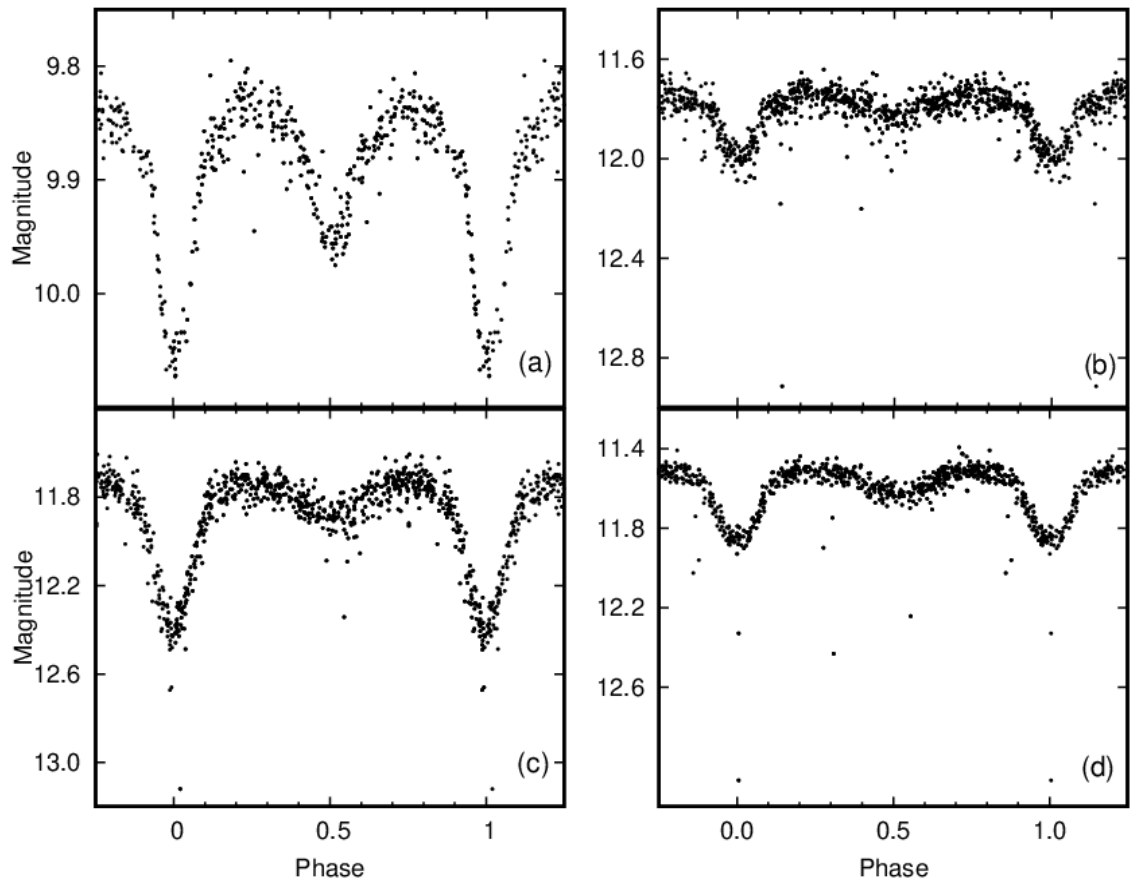

**Figure 1.** The light curve of the system (a) A070, (b) A073, (c) A143, (d) A163 provided by ASAS webpage.

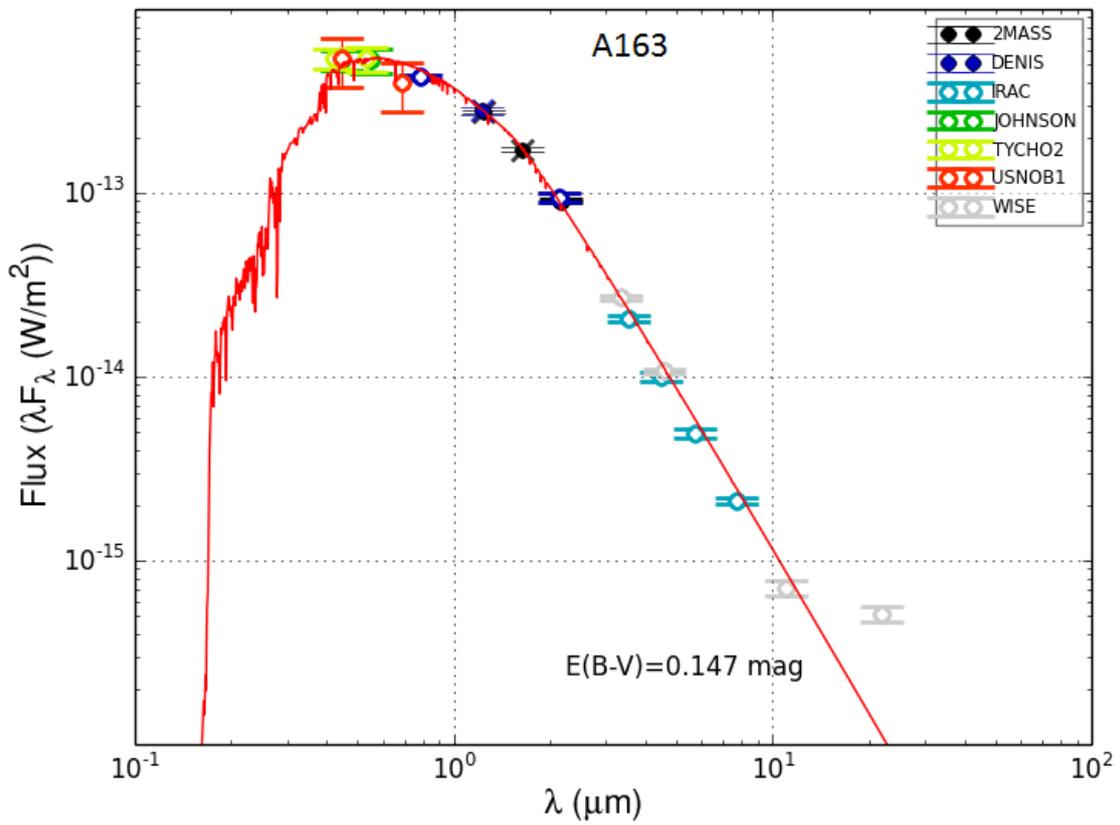

**Figure 2.** The Spectral Energy Distribution (SED) of A163. The legend on the upper right lists the catalogues where the photometric data were taken. The solid (red in coloured edition) line indicates the reddened fit using the Kurucz model.

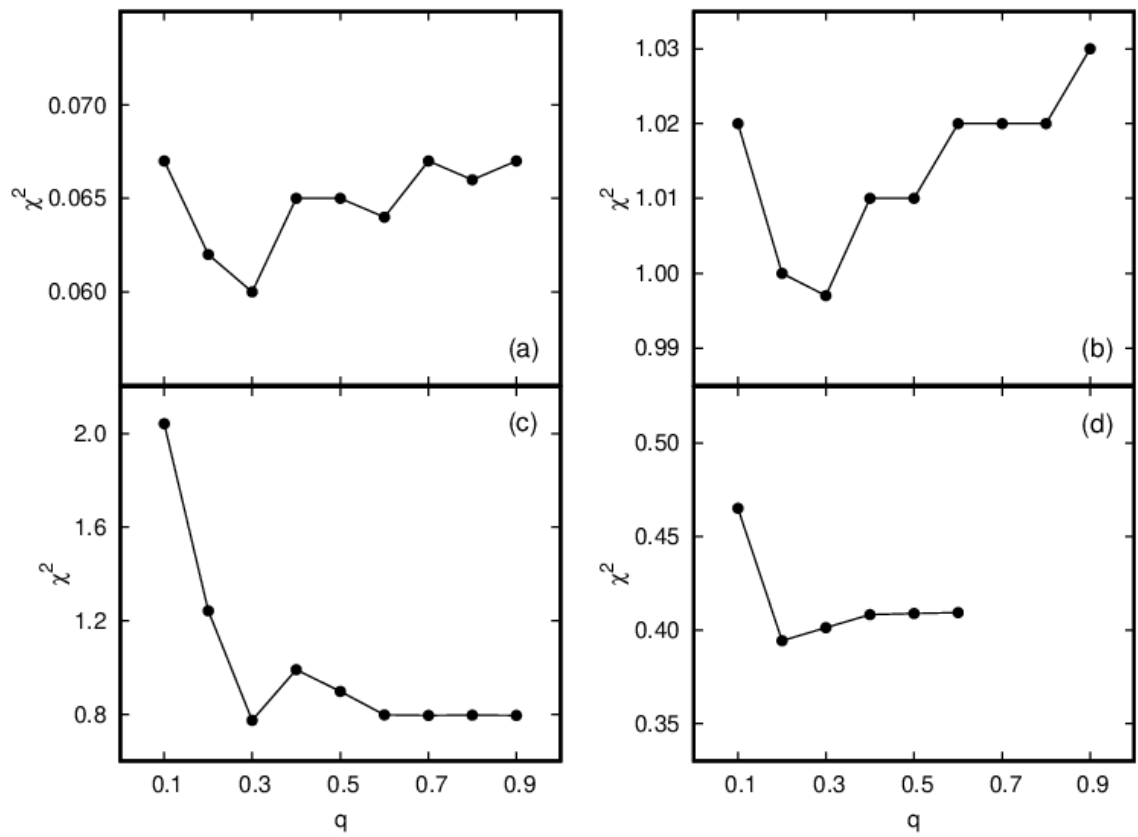

**Figure 3.** The results of the q-search for (a) A070, (b) A073, (c) A143 and (d) A163. The code gives unphysical results for the values higher than 0.6 for A163.

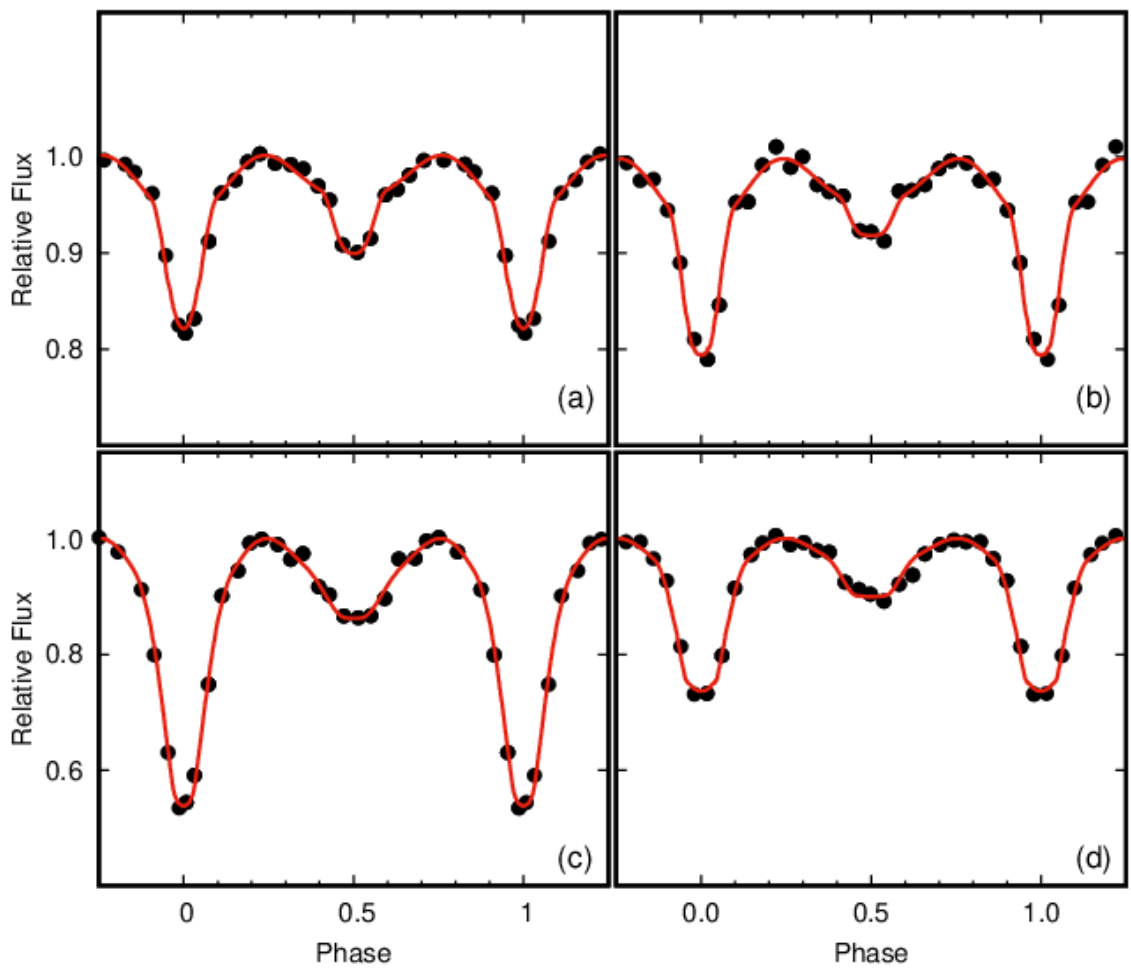

**Figure 4.** The light curves showing the agreement between observations (dots) and the analyses (lines) for (a) A070, (b) A073, (c) A143, (d) A163.

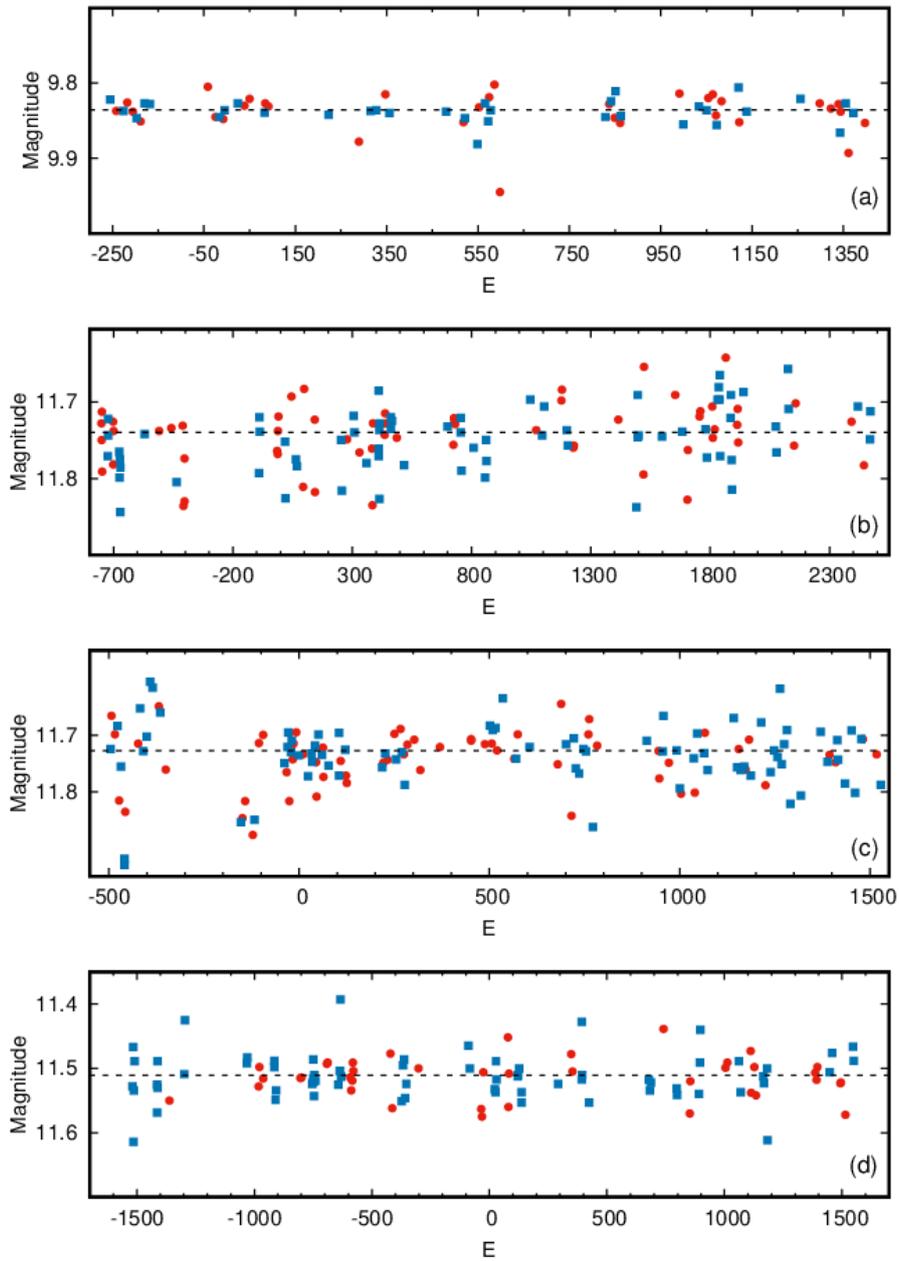

**Figure 5.** The plots of the fluxes at maximum phases versus cycle numbers (E) for (a) A070, (b) A073, (c) A143, (d) A163. The circles (red in coloured edition) and squares (dark blue) refer the first and second quadrature data, respectively. The dashed lines intersect the magnitude values at $9^m.83$ for A070, $11^m.73$ for A073 and A143, $11^m.51$ for A163. Remarkable variations on the data similar to characteristic of O'Connell effect are not observed.

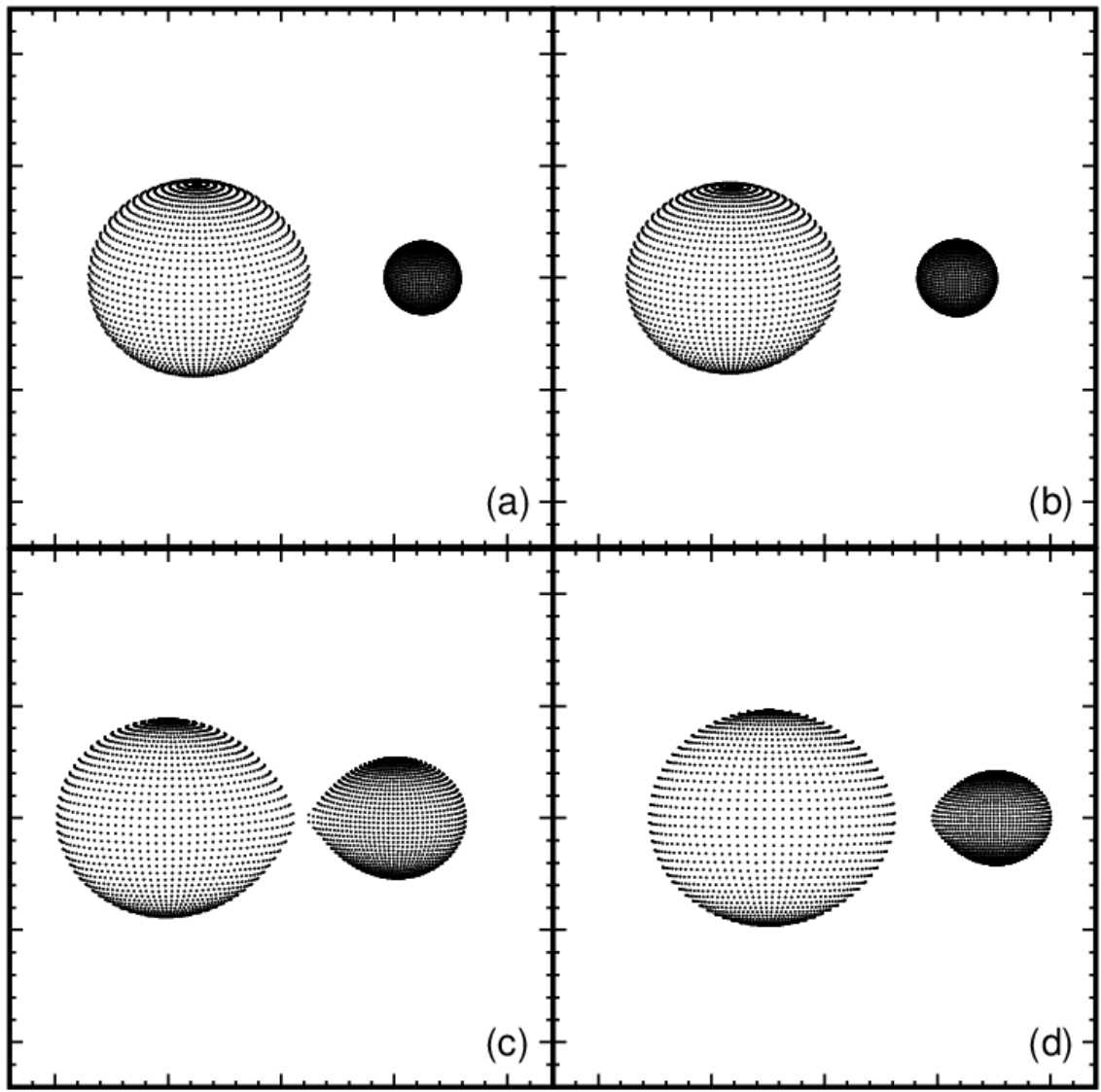

**Figure 6.** The geometry of the systems for (a) A070, (b) A073, (c) A143, (d) A163 at phase Φ=0.25.

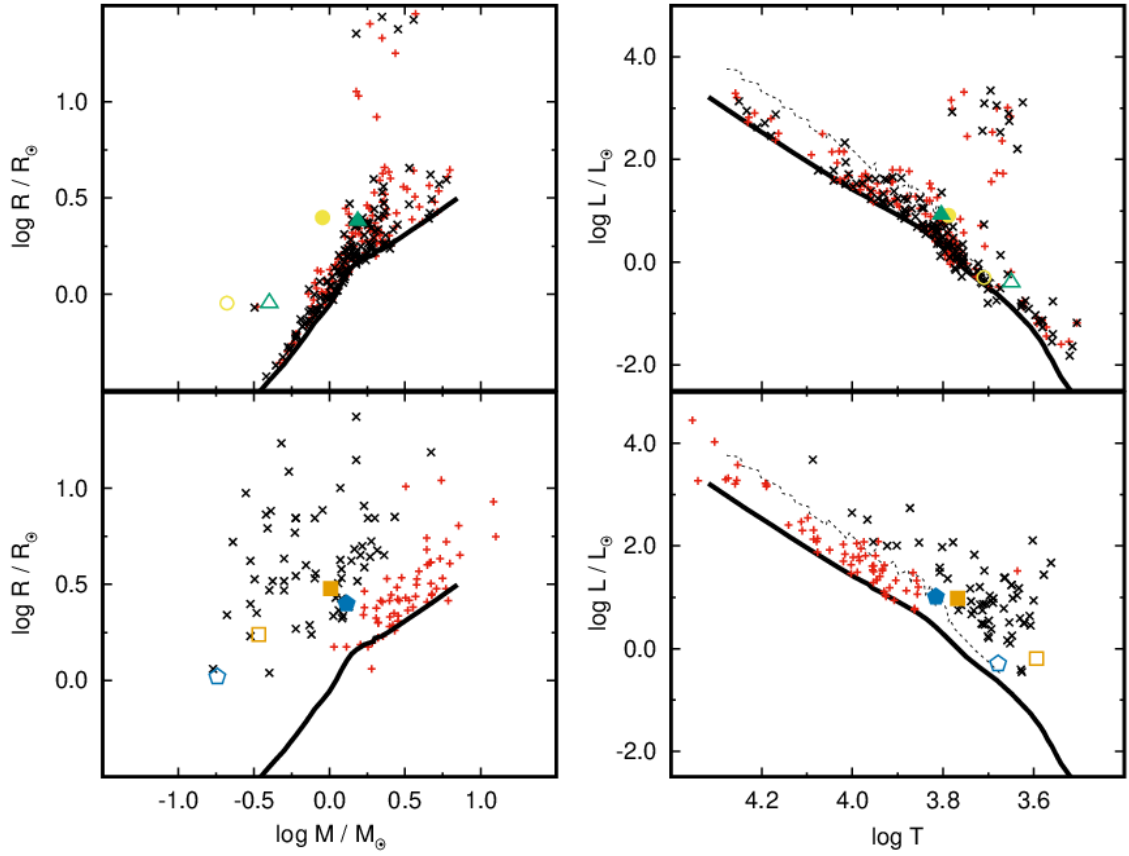

**Figure 7.** The positions of the components of the systems on the mass-radius plane (left) and the Hertzsprung-Russell diagram (right). The plus and cross symbols refer to the primary and secondary components of detached and Algol type binaries whose data taken from [24] and [25]. Circles (yellow in coloured edition), up triangles (green), squares (brown) and pentagons (dark blue) indicate the components of A070, A073, A143, A163, respectively. Filled symbols illustrate the primary components while the open ones stand for the secondaries. Black solid and dotted lines represent the Zero Age Main Sequence (ZAMS) and Terminal Age Main Sequence (TAMS) whose data are taken from [29].

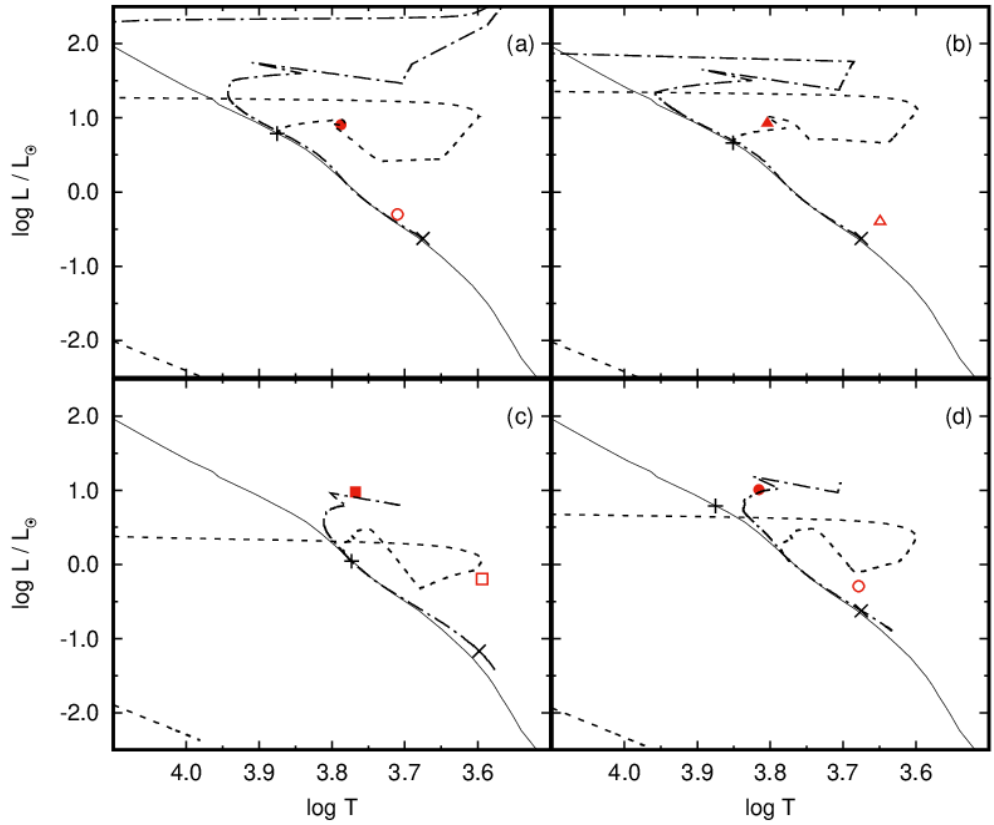

**Figure 8.** The location of the targets on H-R diagram with the evolutionary tracks of the components of a binary system having initial masses $M_{1i} = 1.6\ M_\odot$, $M_{2i} = 0.8\ M_\odot$, initial period $P_i=2.4$ days and the initial eccentricity $e_i=0.6$ for A070 (a), $M_{1i} = 1.5\ M_\odot$, $M_{2i} = 0.8\ M_\odot$, $P_i=6.0$ days and $e_i=0.8$ for A073 (b), $M_{1i} = 1.1\ M_\odot$, $M_{2i} = 0.5\ M_\odot$, $P_i=6.8$ days and $e_i=0.8$ for A143 (c) and $M_{1i} = 1.1\ M_\odot$, $M_{2i} = 0.7\ M_\odot$, $P_i=2.6$ days and $e_i=0.6$ for A163 (d). Filled and open symbols represent the primary and secondary components of the systems, respectively. Plus and cross signs stand for the location of the primary and secondary component of the model in their zero ages. The dashed and dot-dashed lines refer to the evolutionary tracks of primary (more massive) and secondary (less massive) components of the model, respectively. The evolutionary tracks are calculated using BSE code [27, 28]. The solid line represents the Zero Age Main Sequences whose data are taken from [29].